\documentclass[runningheads]{llncs}
\usepackage[T1]{fontenc}
\usepackage{graphicx}
\usepackage{hyperref}
\usepackage{subcaption}
\usepackage{enumitem}
\usepackage{xcolor}
\usepackage{relsize}

\usepackage{caption}
\begin{document}
\title{Aesthetics of Robot-Mediated Applied Drama:\\A Case Study on REMind}
%
%
\author{Elaheh Sanoubari\inst{1} \and
Alicia Pan\inst{2}\and
Keith Rebello\inst{2} \and
Neil Fernandes\inst{2} \and Andrew Houston\inst{3} \and Kerstin Dautenhahn \inst{2}}
\authorrunning{E. Sanoubari et al.}
%
\institute{Systems Design Engineering, University of Waterloo \and Electrical \& Computer Engineering, University of Waterloo \and Communication Arts, University of Waterloo
\\\email{\{esanouba, a29pan, k2rebell, n24ferna, houston, kdautenh\}@uwaterloo.ca}}
\maketitle              
\begin{abstract}
Social robots are increasingly used in education, but most applications cast them as tutors offering explanation-based instruction. We explore an alternative: \textit{Robot-Mediated Applied Drama} (RMAD), in which robots function as life-like puppets in interactive dramatic experiences designed to support reflection and social-emotional learning. This paper presents \textit{REMind}, an anti-bullying robot role-play game that helps children rehearse bystander intervention and peer support. We focus on a central design challenge in RMAD: how to make robot drama emotionally and aesthetically engaging despite the limited expressive capacities of current robotic platforms. Through the development of \textit{REMind}, we show how performing arts expertise informed this process, and argue that the aesthetics of robot drama arise from the coordinated design of the wider experience, not from robot expressivity alone.

\end{abstract}
\section{Introduction}
The recent developments in generative AI has endowed social robots with new capabilities that has made them a rapidly emerging technology.
According to a recent survey of longitudinal deployments, education has been the most popular application area for social robots over the past two decades \cite{matheus2025long}. In the domain of education, physical robots outperform virtual agents: they can improve children’s learning and impact behavioral changes more substantially \cite{belpaeme2018social}, and learners consistently demonstrate higher trust, enjoyment and compliance when interacting with them \cite{breazeal2016social}.
A review of social robots in education attributes this effectiveness to physical embodiment (which traditional learning technologies lack), and concludes that robots are “a natural choice when the material to be taught requires direct physical manipulation of the world”, such as handwriting or throwing a ball \cite{belpaeme2018social}. However, in the vast majority of applications in education, social robots are used as a stand-in for a tutor or teacher, whose primary role is to provide direct curriculum support (e.g., teach math, or a second language) via explanation-based didactic pedagogy \cite{belpaeme2018social,matheus2025long}. In rare cases where a robot is used to teach a physical skill such as hand-writing \cite{chandra2020children}, although the results are promising, the robot does not physically hold a pen to demonstrate handwriting, due to the limited manipulation abilities of its mechanical hands. The field is yet to tap into the full potential of social robots in education.

What makes robots unique is that they are anthropomorphized more than other forms of technology; and physical embodiment is a critical factor in this process. 
Evolutionary psychologists suggests that the human visual system is biologically tuned to prioritize the detection of animals and people, which is rooted in our \textit{ancestral priorities} \cite{new2007category}. The autonomous movement of robots as embodied agents in our space activates the ``life detector'' in human brains, and we are hardwired to attribute a \textbf{state of mind} to robots and perceive them \textbf{as social actors}. Hence, their greatest potential for education may be their capacity to be used as artificial social actors akin to puppets, which have long been used by therapists and educators to support children’s social-emotional learning—except, robots afford multimodal autonomous behavior, interactivity, and programmability, which puppets lack.
Tapping into the full potential of social robots in education means going beyond the pedagogical paradigm of using robots as (less capable) stand-ins for teachers who use explanation to deliver content to learners; and instead, shifting to a drama-based pedagogy.

We propose \textbf{Robot-Mediated Applied Drama (RMAD)} \cite{sanoubari2022robot-mediated}: a pedagogical framework in which robots are used as life-like puppets, and learning is organized as interactive drama practices (such as role-play, and improvisation) in make-believe simulations of real-life social situations, in order to promote inner reflection. Our prior work explores the theoretical foundations of RMAD, and highlights that the focus is on the process of Applied Drama and the participant reflection it elicits, rather than staging a Theater performance \cite{sanoubari2022robot-mediated}.

Facilitating meaningful RMAD experiences presents unique challenges. While robots can be used as actors, they have significant constraints for creating and maintaining the aesthetic experience that is central to drama. This manuscript presents a case study on facilitating RMAD: \textit{REMind}, an anti-bullying robot role-play game named \textit{REMind} that is designed to help children rehearse bystander intervention and peer support.
Drawing on the development of \textit{REMind} through iterative technical rehearsals with a theater director, we examine how robot-mediated dramatic interaction can be designed to remain aesthetically engaging despite the constraints of current robotic platforms.

This paper makes three contributions. First, we present RMAD as a performing-arts-informed approach to HRI in which robots function as life-like puppets in interactive dramatic learning experiences. Second, we show how feedback and technical rehearsals with performing arts expertise can serve as a method for facilitating, developing and refining robot-mediated interaction. Third, we argue that aesthetic engagement in robot drama is distributed across a broader performance ecology---including facilitation, scenography, props, timing, and participant action---rather than residing in robot expressivity alone.

\section{Background: Understanding Aesthetics}
Drama Education engages the learners’ mind, body, and emotions in a holistic way that is in many respects \textit{superior} to explanation-based pedagogy~\cite{decoursey2018embodied}. Participants of drama are not passive listeners, they \textit{do} embodied things. 
DeCoursey describes drama as ``a special way of experiencing the world'', and emphasizes that Aesthetics has a \textit{functional} place in drama education~\cite{decoursey2018embodied}. The `Aesthetic Experience' is conceptualized as a moment involving three components: intense attention, emotional response, and pleasure in the process ~\cite{decoursey2018embodied}. The centrality of emotion in this process puts natural constraints on the capacity of social robots to create and maintain the aesthetic experience of drama. 
A real actor can naturally grasp the intended emotion for the character upon reading the script (even if it is not explicitly stated), and believably pretend to experience it in a performance; triggering the audience to \textit{feel with} them \footnote{A process referred to as Double-Noesis ~\cite{decoursey2018embodied}}. The state of art social robots have limited affective, expressive and perceptual capacities which significantly constraints this process.

Aesthetic is also an important discussion in game design literature. Schell’s \textit{Elemental Tetrad} of game design~\cite{schell2008art} breaks games down into four interdependent elements, one of which are the aesthetic elements which determine how the game looks, sounds, and feels. Aesthetics are the most visible layer of the experience that constitute the player's immediate sensory engagement, and has the most  most direct relationship with the player’s experience~\cite{sellers2017advanced,schell2008art}.
In video games, aesthetics typically encompass audio elements (such as music and sound effects), and visual elements (such as graphics and art style)~\cite{schell2008art}. Sellers also connects aesthetics to a game’s use of the player's ``interactivity budget,'' suggesting that a fast-paced game “can afford to be assaulting the player’s senses with feedback”~\cite{sellers2017advanced}. This idea aligns with the notion of “juiciness” in interaction design: a juicy game is one that incorporates various layers of audio-visual embellishments in its interface feedback~\cite{hicks2019juicy}, such that just a little bit of interaction gives the player a continuous flow of sensory rewards~\cite{schell2008art}.

Game designers often use aesthetic elements to set context for the emotions they want players to feel, for example by using colors and music to prime players for emotions such as fear, victory, or hope~\cite{sellers2017advanced}. 
This focus on the player's desired emotional state is formally structured by the Mechanics-Dynamics-Aesthetics (MDA) framework~\cite{hunicke2004mda}, which defines Aesthetics as encompassing the ``desirable emotional responses evoked in the player,'' and provides a taxonomy of 8 different kinds of pleasures in video games. Prior work has extended this taxonomy to robot games~\cite{sanoubari2024makes}.
While MDA equates game aesthetics with “fun,” some game designers differentiate between the two. Koster treats ``aesthetic appreciation'' as one of the factors contributing to enjoyment of games~\cite{koster2013theory}. He argues that if the core game is inherently fun, the aesthetics serve to focus and magnify that fun; otherwise, no amount of aesthetic ``dressing'' will make it enjoyable~\cite{koster2013theory}. 

Irrespective of the medium of communication, Schell argues, the “poetry of presentation” can make experiences more compelling~\cite{schell2008art}.
This perspective resonates with the Human-Computer Interaction (HCI) literature, particularly Norman’s framework of Emotional Design~\cite{norman2005emotional}, which emphasizes the cognitive and behavioral significance of emotional responses. Norman asserts that the emotional aspect of interacting with an artifact is often more crucial to a its success than its functional elements, arguing that “attractive things work better”~\cite{norman2005emotional}.

This paper builds on this interdisciplinary understanding of to reflect on the Aesthetic Experience of \textit{REMind}, and its design implications for RMAD.

\section{Case Study: RMAD in an Anti-Bullying Intervention}
Bullying is a `group process' shaped not only by bullies and victims, but also by bystanders, whose silence or encouragement influence the outcome.

Although many peers disapprove of bullying, they often do not intervene because of social-emotional barriers such as fear, uncertainty, or lack of confidence. Traditional anti-bullying programs provide learners with \textit{knowledge} rather than \textit{skills}; for example, by teaching children what bullying is, or giving them a `script' for intervention~\cite{sanoubari2022designing}), which rarely translates into situated competence needed to \textit{actually} intervene.
This critical gap suggests that anti-bullying interventions should provide situated opportunities for learners to practice how to respond.

\begin{figure}
    \centering
    \includegraphics[width=0.6\linewidth]{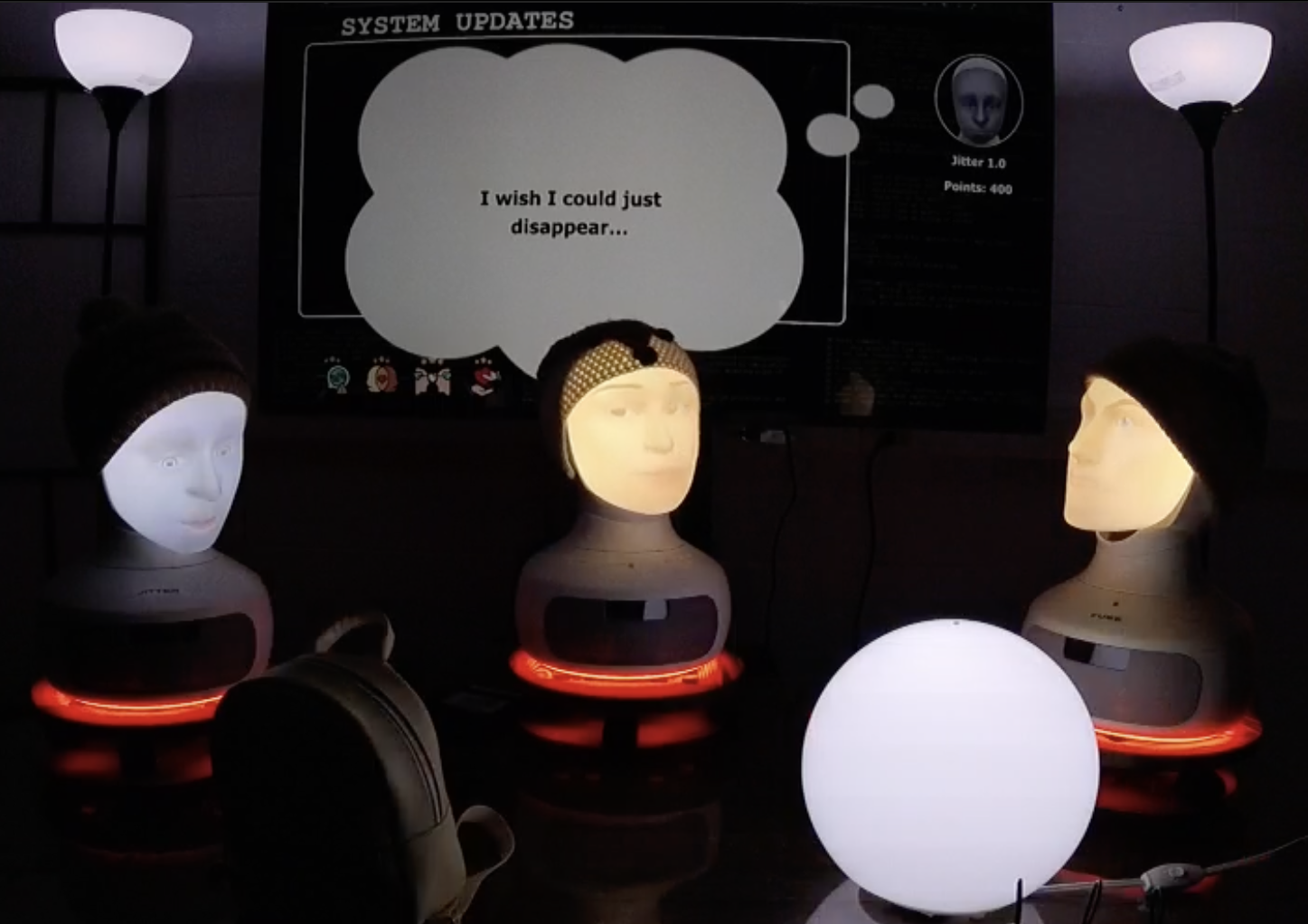}
    \caption{\textit{REMind} engages children in role-play with three Furhat robots.}
    \label{fig:REMindGameView}
\end{figure}

Applying RMAD to this challenge, we developed an anti-bullying robot role-play game called \textit{REMind} (see Fig. \ref{fig:REMindGameView}). In this game, a child player is tasked with mentoring a robotic avatar to stand up to bullies. The player is guided through a step-wise make-believe process of `trouble-shooting' a mysterious ``glitch'' caused by robot's past failure to intervene in a peer bullying incident. They help avatar reflect on peer bullying, develop empathy, take perspective, and finally, intervene.

\textit{REMind} draws on Augusto Boal’s approach to participatory drama, which treats theatre not as entertainment, but as a rehearsal space for social change, which can empower ordinary people to stand up to social injustice \cite{boal2000theater}. In Boal's Forum Theatre, the play is performed twice——The second time, audience members are invited to become ``spect-actors'': active participants who can interrupt the performance at a key moment, and propose alternative actions for the protagonist to lead the scene toward a different conclusion~\cite{boal2000theater}.
Similarly, \textit{REMind} allows children to first watch a live drama by three social robots in which one robot bullies another, and avatar remains a passive bystander. Then, players practice spect-actorship by using `Puppet Mode:' acting out an alternative response on behalf of avatar, and streaming a recording of it onto the robot in a replay of the bullying scene. In this way, the player can see their idea come to life, and observe how the narrative unfolds as a consequence of their intervention.

\subsection{Iterative Design Journey}
The iterative design journey of \textit{REMind} involved multiple stakeholders, including children, educators, and subject-matter experts. In interviews with local elementary teachers (N=13) we learned that interventions must move beyond outdated, adult-centric scenarios~\cite{sanoubari2022designing}. 
This led to a narrative co-design study (N=22) where children generated the core characters and storyboards~\cite{sanoubari2021robots}.
Applying game-based learning principles, we conducted focus groups with children (N=15) to elicit design ideas for making \textit{REMind} `fun'; and performed a Framework Analysis by using the Aesthetics category of the MDA framework as a lens, which yielded 28 elements (such as use of props, and environmental storytelling) that were used for refine interactive elements and game mechanics of REMind~\cite{sanoubari2024makes}.
We then consulted two expert psychologists who had led the development of KiVa~\cite{salmivalli2012kiva} (a widely adopted evidence-based anti-bullying program) to align game mechanics and narrative elements with the pedagogical objectives and metrics. To translate these designs into a live RMAD experience, we establisehd a facilitation protocol in consultation with audience research experts.
Finally, \textit{REMind} was iteratively prototyped in Technical Rehearsal sessions with a theater director, which shaped the designs into a 2-hour experience. It was evaluated in play-testing trials with children (N=18), which yielded a significant increase in self-efficacy for intervention, and generated evidence of learning in all pre-determined learning goals, and will be discussed in future manuscripts.

\subsection{Facilitation: Learning from Audience Research Experts}
Since a primary objective of RMAD is to elicit participant reflection, facilitation becomes as important as the technical design. Initially, we planned to facilitate \textit{REMind} using the Wizard of Oz (WoZ) technique (i.e., via a hidden operator)~\cite{riek2012wizard}. Experts on Spectatorship and Audience Research critiqued the initial facilitation structure and suggested assigning a ``Joker'' to mediate the experience. The Joker is an essential facilitator role in Forum Theatre~\cite{boal2000theater}, who uses improvisation to meditate between the actors and the audience, guiding reflection, and encouraging spect-actorship. Importantly, the Joker does not impose solutions but helps create a space for playful problem-solving, keeping the process adaptive, participatory, and focused on exploration rather than prescription.

Learning from this, \textit{REMind} prototypes were semi-autonomous and relied on a two-person facilitation protocol: a `Wizard' (hidden WoZ operator) and `Joker' (in-room facilitator). The Wizard controlled narrative progression and made simple branching decisions based on player’s response (e.g., game choices). A researcher took the role of the Joker by adopting the persona of a 'programmer' in the drama. 
Their role included improvising dialogue to handle unscripted questions, and troubleshooting technical issues without breaking the child's narrative immersion. During specific modules the Joker posed open-ended questions to deepen reflection (e.g., \textit{``How would you explain `Sadness' to a robot?''}). While the Joker's actions were responsive and adaptive, they followed a semi-structured facilitation protocol. In this context, the Joker does not introduce variability, but rather provided a consistent structure for learners' interpretation and meaning-making, which are core goals of RMAD.

\subsection{Technology}
\label{subsection:Technology}
\textit{REMind} uses three Furhat robots, GUI, immersive audio-visual cues, and physical props. While the complete documentation of the technological implementation is beyond the scope of this paper, this section will provide a brief overview.

\subsubsection{Setup}
\begin{figure}
    \centering
    \includegraphics[width=0.6\linewidth]{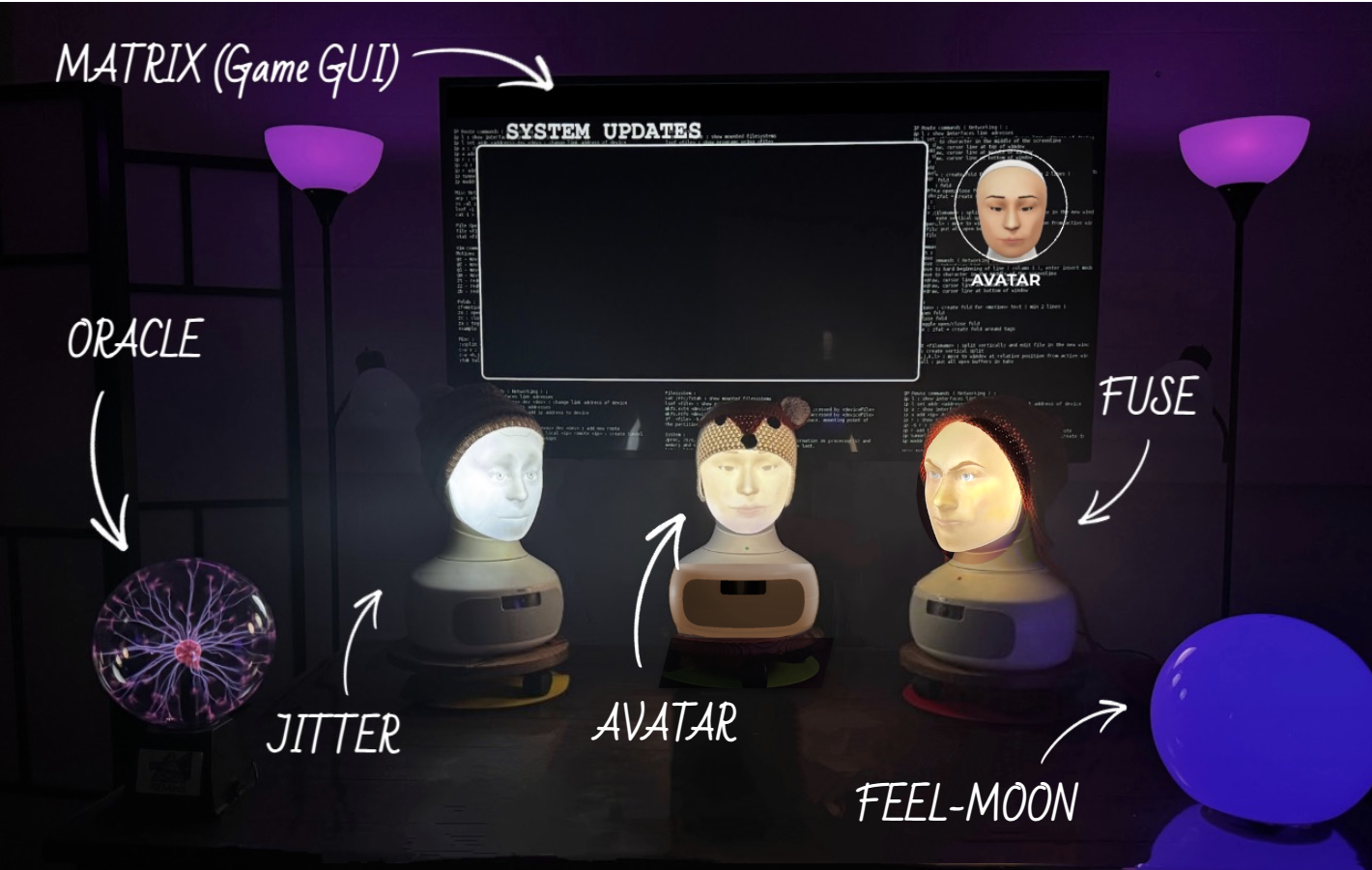}
    \caption{Storytelling elements in the \textit{REMind} game world.}
    \label{KERNELSetup}
\end{figure}

\textit{REMind} takes place in a room named \textit{KERNEL}, a robotic lab at the University designed to resemble a dimly lit studio apartment. The central interaction space features a dining table where the three Furhat robots (\textit{AVATAR}, \textit{FUSE}, and \textit{JITTER}) are arranged in a triangle. Directly behind the table is a large TV named \textit{MATRIX}, which serves as the primary game GUI. This area also houses key interactive props used in the role-play, including a a plasma light named \textit{ORACLE} and a globe-shaped lamp named \textit{FEELMOON} (see Fig. \ref{KERNELSetup}).

\subsubsection{Robotic Platform}
Furhat~\cite{al2012furhat} is a tabletop robot with a 3-D printed face, capable of projecting facial expressions, head movements, and speech. Robot speech was synthesized using Microsoft Azure Text-to-Speech (TTS) engine. 

\subsubsection{Rapid Prototyping Toolkit}
Scripting a synchronous multi-robot, multi-modal interactive experience presents unique engineering challenges. Unlike video games where all elements are virtual, or live theater where actors adapt on the fly, robots are embodied agents with limited capacity for improv. They cannot dynamically interpret unscripted interaction, autonomously synchronize, or improvise social-emotional nuances.

\begin{figure}
    \centering
    \includegraphics[width=1\linewidth]{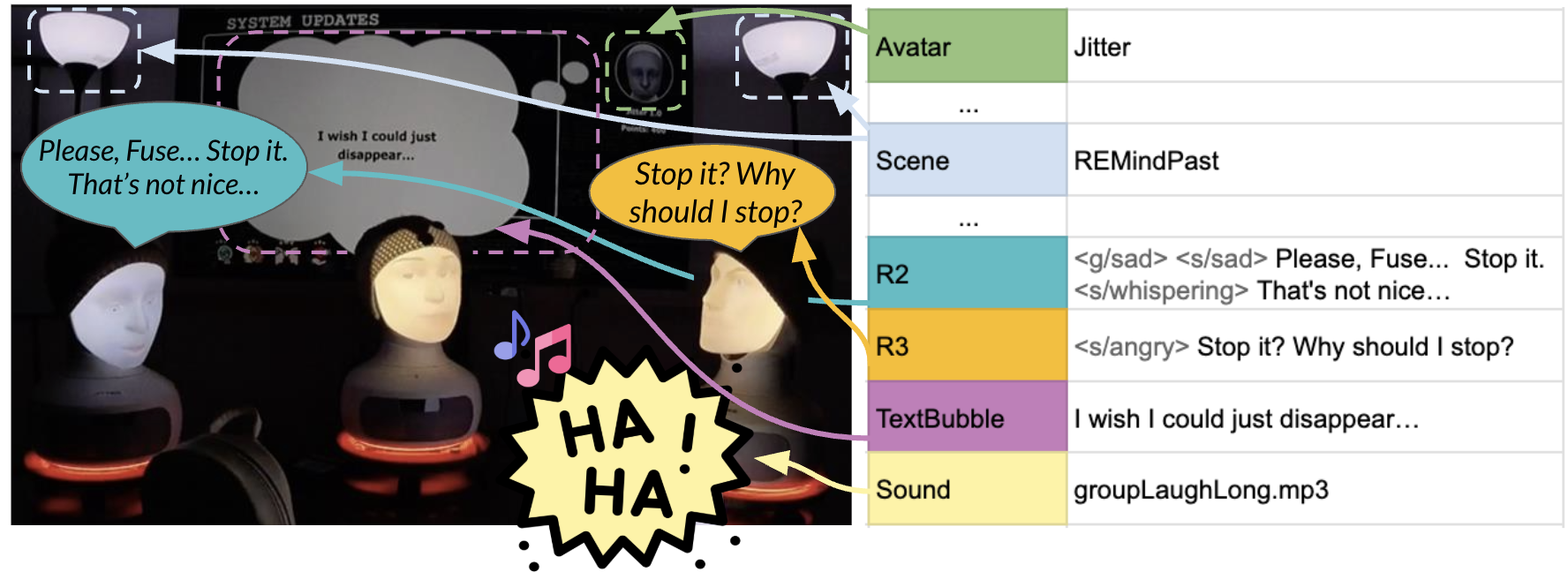}
    \caption{Example translation from \textit{StorySync} script into a live scene in \textit{REMind}.}
    \label{fig:StorySyncToScene}
\end{figure}

There are existing tools for scripting interactive narratives for video games (e.g., Articy Draft\footnote{\href{https://www.articy.com}{articy.com}}), and synchronizing theatrical cues (e.g., QLab\footnote{\href{https://qlab.app/}{qlab.app}}). However, there is no publicly available authoring toolkit that bridges multi-modal ambient cues (à la theatre), narrative branching logic (à la games), and robot choreography—enabling designers to rapidly prototype interactive and embodied storytelling systems.
To rapidly prototype \textit{REMind}, we developed StorySync: a custom-made spreadsheet-based authoring toolkit that allowed us to synchronize the entire digital narrative system by using avatar as a centralized server. Each row in the spreadsheet defines a scripted event that can be triggered either sequentially, or via a human operator in the loop. Fig. \ref{fig:StorySyncToScene} illustrates an example of \textit{StorySync} driving a live scene in \textit{REMind}. 
\vspace{-10pt}

\subsubsection{Gesture Capture}
During Puppet Mode, we used the Live Link Face iOS application by Unreal Engine~\cite{epicgames_livelinkface} to capture high-fidelity facial motion data. This application allows exporting facial expressions in CSV format by using Apple’s ARKit~\cite{apple2025arkit} facial parameters. Using the Furhat Gesture Capture tool~\cite{furhat2025gesturecapture}, we converted the facial parameters into Furhat-compatible \textit{.json} gesture files. Finally, we paired this gesture with the player's audio from the original recording.

\subsection{Player Experience Overview}
The \textit{REMind} gameplay experience is structured around a five-phase interactive narrative. Full account of the narrative design, including game props and how they are used as narrative mechanics, will be discussed in future publications.

\subsubsection{Phase 1: Entering the Magic Circle}
The experience begins with the player customizing the robotic avatar by selecting its face, voice, name (using a GUI),and physical accessories.
To start the game, the player places their hand on \textit{ORACLE} and speaks out the command, \textit{`Let the Adventure Begin!'}. This interaction causes the room's lights to brighten, a startup video to play, and the \textit{AVATAR} robot to wake up. \textit{AVATAR} immediately complains about a ``glitch'' that is causing robots to malfunction (e.g. acting selfishly, forgetting names of friends), explaining that it needs help from the child because it ``lacks emotional intelligence''.

\subsubsection{Phase 2: Simulating the Memory \& Fixing the Bully-Detector}

To investigate the glitch, the player finds a floppy disk prop and inserts it into a reader to simulate \textit{AVATAR}'s past memory, which triggers the other two robots to wake up. In the memory, one robot (FUSE) bullies another robot (\textit{JITTER}) while \textit{AVATAR} passively observes. \textit{AVATAR} claims it did not intervene because it thought the bullying was ``just a joke''. The GUI flags that \textit{AVATAR}'s `\textit{Bully-Detector}' is broken. The player finds the `\textit{Bully-Detector}' (a custom-made interactive prop, see \cite{sanoubari2024makes}) exploring it to teach the robot how to identify bullying.

\subsubsection{Phase 3: Debug Mode (Empathy-Training \& Perspective-Taking)}
The player activates \textit{AVATAR}'s `Debug Mode', which displays missing social information on the \textit{MATRIX} (e.g. `\textit{JITTER}'s feelings). \textit{AVATAR} explains that because robots cannot feel emotions, it needs help from the player to interpret them. Using \textit{FEELMOON}, paired with ambient music, \textit{AVATAR} displays two emotions ``detected'' in \textit{JITTER}. The first display is a dim, pulsating blue light and melancholic piano music, representing sadness. The second is a fast-pulsating purple light paired with a tense, and somewhat ominous music and heartbeat sound, representing fear. The player identifies and explains the emotions, and guesses the underlying thoughts of the victim and the motivations of the bully.

\subsubsection{Phase 4: Puppet Mode (Robot-Miedated Spect-Actorship)}
\textit{MATRIX} displays two intervention strategies: `\textit{Comfort JITTER}' (for 500 points) or ` \textit{Be Firm with FUSE}' (for 1000 points). The player selects one and demonstrates it. Their improvised performance is recorded and converted into a Furhat robot gesture. The player re-simulates the bullying scene, and at the appropriate moment, activates \textit{`Puppet Mode'}, which causes \textit{AVATAR} to interrupt the bully, acting out the exact facial expressions and voice provided by the player.

\subsubsection{Phase 5: Outcomes and Resolution}

In the final phase, the player uses game props to explore possible endings of the story from the perspectives of different characters. In doing so, they discover three important outcomes of intervention: the bully may back down or retaliate; even if the bully retaliates, intervention will make the victim feel better and more supported; and finally the defender’s peer status increases. The player and \textit{AVATAR} then uncover the true source of the ``glitch''---a shared misunderstanding of social norms---and symbolically resolve it, allowing \textit{AVATAR} to remember its friends’ names.

\section{Technical Rehearsals with a Theatre Director}
We iteratively prototyped \textit{REMind} during 10 Technical Rehearsal sessions (\~20 hours), spread-out over 3 months. These sessions involved the Wizard, the Joker, and a theater director, as well as a adult pilot tester—typically a colleague unfamiliar with the system who stood in for the child player. During technical rehearsals, we built evolving prototypes of \textit{REMind}, using observations and feedback to refine both the system and the performance structure.

The primary purpose of the technical rehearsals was prototyping: they allowed us to investigate how the technology could support the storytelling, interaction, and dramatic structure of the performance; and conversely, how the developing performance shaped the technical requirements of the system. In theatre practice, technical rehearsal is the stage at which the ``weave of actions'' is tightened~\cite{barba2005secret} —when timing, sequencing, and coordination of actions (performed by the robots, GUI, wizard, and Joker) are refined in relation to the intended meaning of the piece.
In this sense, our rehearsal process was guided by dramaturgy: continually asking if what we are doing means what we think it means.
At the same time, these sessions exposed the practical constraints and affordances of the hardware and software, enabling us to identify where the original design was unworkable, where it needed adaptation, and where unexpected possibilities emerged. Thus, technical rehearsal functioned as a process of negotiating between dramaturgical intent and technological reality.
Beyond prototyping, three design outcomes emerged from technical rehearsals: training the Joker, Wizard--Joker coordination, and affective tuning of the robot drama. The WoZ toolkit continued to evolve to support the changing requirements.

\subsubsection{Training the Joker}
Technical rehearsals were essential for training the Joker as the Forum Theatre facilitator of REMind. Early on, this meant learning the practical boundaries of the system: what the robots could do reliably, where timing was fragile, and how to pace the interactions. Over time, the rehearsals also became a space for developing improvisational skill. For example, through pilot-testing, the Joker learned to address technical issues occurred while staying in character by explaining that the robot was ``glitching,'' which helped with preserving narrative immersion. As the Joker gained experience with the system and pilot-testers, they developed a better sense of when to slow down, when to probe further, and how to adapt their responses while staying within the narrative, pedagogical, and technical constraints of the performance.

\subsubsection{Wizard--Joker Coordination}
Technical rehearsals also helped establish coordination practices between the Wizard and Joker. The pair developed a theater-inspired communication protocol using subtle visual and verbal cues. For example, in modules where the Joker asked follow-up reflective questions, the WoZ interface instructed the wizard to wait until the Joker gave a thumbs-up before advancing. If a problem arose in the room that the Wizard could not see directly (such as the robot gaze being misdirected) the Joker responded in character by placing a hand on the plasma-light prop and saying, ``Hey \textit{ORACLE}, help the robot look at the player.'' The rehearsals helped the pair develop live timing and repair strategies that could be used without breaking the dramatic frame.

\subsubsection{Affective Tuning of the Robot Drama}

A central challenge in \textit{REMind} was making the robot drama feel legible, believable, and emotionally resonant using constrained expressive resources. Technical rehearsals provided the setting in which we repeatedly tested and revised robot gesture, dialogue, vocal prosody, and ambient cues to improve this dramatic legibility. Since this process became a substantial design effort in its own right, we discuss the challenges and strategies for programming emotional subtext in the following section.

\section{Lessons Learned: Programming Emotional Subtext}
\label{programming_subtext}
Stories are meant to be told. To bring a narrative to life, human performers leverage their embodiment, using motion, timing, gestures, and vocal variation to convey meaning. 
In performance and communication, this is referred to as `subtext': the unspoken meaning or emotions behind what a character says in a script, which signals what they are really feeling or thinking, adding depth to their words and actions.
Creating compelling robot drama is challenging because robots have limited emotional expressivity and struggle with intentionally conveying emotional subtext.
This has critical implications, because emotions have a key role in how stories persuade us and foster empathy.

\subsection{Challenges with Facial Expression}
\begin{figure}
    \centering
    \includegraphics[width=0.7\linewidth]{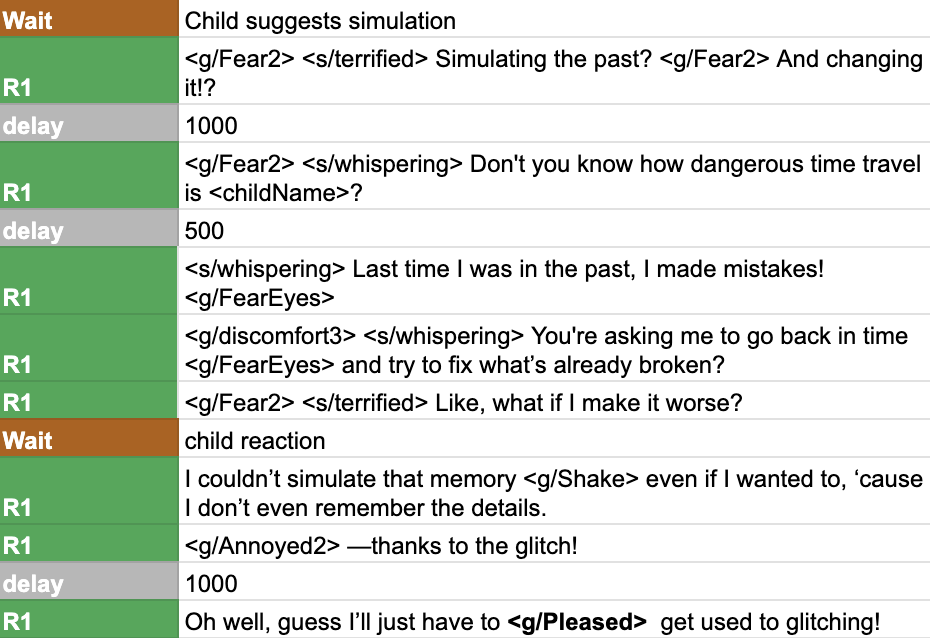}
    \caption{\textit{REMind} script using a gesture labeled `Pleased,' to convey fear.}
    \label{fig:pleased_or_not_pleased_that_is_the_Q}
\end{figure}

Facial expression is a primary channel of conveying emotional subtext in humans.
Our decision to use the Furhat robots was largely because unlike many robotic platforms, they afford facial expressions.
While the Furhat platform offers a library of basic facial expressions (aka. gestures), we encountered several challenges in using them to create believable emotional subtext. 
First, facial expressions are highly contextual, which makes it challenging to script robot drama using a library of pre-defined gestures. Even gestures with seemingly straight-forward labels are inherently vague. For example, a `smile' can look genuine, sarcastic, or apologetic. Similarly, a pre-defined gesture labeled `thoughtful' from a library may communicate concentration, confusion, or mild sadness; making it difficult for designers to discern the expression based on the tag in order to use it for precise scripting of emotional expressions. Therefore, the only reliable way to understand whether the gesture is an appropriate match to the dialogue context is trial and error. Sometimes this resulted in counter-intuitive scripting choices, such as the \textit{StorySync} screenshot in Fig. \ref{fig:pleased_or_not_pleased_that_is_the_Q} illustrating that a gesture semantically labeled `Pleased' was used to communicate avatar shrugging off its terror in an attempt to ``act cool''. This shows that how well a robot gesture works in tandem with the script supersedes the label it is annotated with.

Second, the reliance on discrete, canned emotional labels makes blending gestures to create nuanced emotional expressions infeasible; for instance, one cannot easily overlay gestures to script a robot behavior that looks ``sad but trying to seem happy,'' or ``confused and mildly irritated.'' Finally, timing, pacing and transition between gestures is difficult. Unlike human actors, a robot actor may jump from one gesture to another abruptly, or fail to dynamically match the length of a gesture to the dialogue; which can manifest as a jarring performance.

\subsection{Challenges with Affective Prosody}
Parallel challenges arose with conveying affect using TTS. Azure TTS offers Speech Synthesis Markup Language (SSML) tags for customizing the `style' of utterances (namely: angry, cheerful, excited, friendly, hopeful, sad, shouting, terrified, unfriendly, and whispering). 
While these tags are useful for adding prosodic variability to synthesized speech, the models often produce a stereotypical, or even exaggerated, version of the emotion. For example, using the excited `excited' tag for a line intended to convey subtle encouragement (e.g., \textit{``Sounds like a good idea, you should try it.''}) results an overly enthusiastic high-pitched rapid delivery that sounds affectively incongruent in the dramatic context.

Also, most TTS engines lack SSML tags for emotive vocalizations, such as laughing, crying, or sighing. While they are offered by some TTS engines (e.g., Acapela Group's `Voice Smileys'), using them in a script is challenging—just because a tag produces a generic \textit{`haha'} does not mean it is the right kind of \textit{`haha'}. Nuanced affective distinctions such as a joyful chuckle, a sarcastic snort, or a sinister cackle remain beyond the expressive granularity of state of art TTS.

Additionally, TTS systems struggle with dynamic emotional shifts. Human speech naturally includes irregular pauses, breaths, or sudden tempo changes to convey subtext. Programmatically replicating prosodic variability in synthesized speech is labor-intensive, and can make the speech sound disjointed. Such limitations mean that a script written to convey a specific emotional tone often sounds completely different when performed by the robot, rendering the traditional approach of ``script first, rehearsal later'' impractical for RMAD.

\subsection{Affective Tuning Strategies}
In technical rehearsals, we identified and iteratively refined snippets of robot perfkormance that seemed affectively incongruent. Affective tuning was labor-intensive; so, we reserved highly granular adjustments for the most emotional parts of the story. This process yielded several strategies for adding subtext.

\subsubsection{Refining Gestures}
To address the ambiguity of the gesture library, the designers tested all the available gestures and annotated them situationally to account for the dramatic context (e.g., distinguishing an \textit{``I'm figuring out the murderer''} thinking gesture from a \textit{``Trying to remember where I parked''} thinking gesture). This provided a more reliable, context-specific vocabulary, and made it easier to remember the content of gestures and quickly adjust the script.  
For moments requiring more nuanced emotional expression, we also moved beyond the predefined gesture library. 
For instance, to convey despair, we created a custom expression in which the robot shook its head, sighed, and looked downward. To do so, we first observed the robot performing the dialogue, then recorded ourselves enacting the intended facial motion, and converted that performance into a custom facial expression executable on the robot using Gesture Capture.

\subsubsection{Dialogue Adjustment}
Many lines in the initial script were too prosodically complex for TTS to perform correctly (e.g., \textit{``Oh wow, You *actually* did it!''}).
To mitigate this, we shifted the script's reliance, wherever possible, from \textit{showing} emotion through prosody and gesture, to \textit{telling} emotion through dialogue. This often resulted in scenes where the robot explicitly verbalized \textit{shifts in feelings} to provide appropriate narrative feedback. While this approach violated the `show, not tell' golden rule of storytelling, it was a necessary design trade-off.
For example, after the empathy training module (where the child taught \textit{AVATAR} about the victim's feelings), the robot verbalized its new-found emotional understanding with this monologue: \textit{``I was there too, but I didn’t feel afraid, I didn’t feel sad. That’s why I thought it wasn’t a big deal! But... If I were JITTER? ... I would have wanted someone to notice! I would have wanted someone to say something!''}

\subsubsection{Prosodic Refinement}
\begin{figure}
    \centering
    \includegraphics[width=0.75\linewidth]{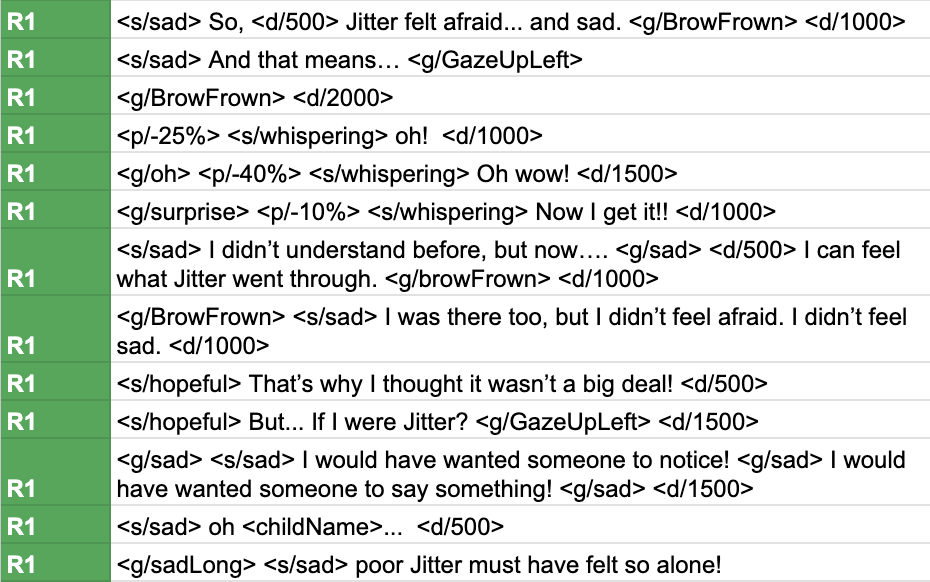}
    \caption{Script of robot's reaction to `empathy training' was affectively tuned by adding gestures ($g/$), and pauses ($d/$), tweaking speech rate ($p/$), and style ($s/$).}
    \label{fig:ssml_finetuning}
\end{figure}

For the most emotional bits of the story, we incorporated a highly iterative process for nuanced affective tuning the vocal delivery. The group would watch a specific short scene from the robot drama, and manually refine the dialogues by experimenting with SSML styles, and tweaking prosodic attributes such as pauses, volume, speech-rate and pitch, until the delivery was compelling (see an example in Fig. \ref{fig:ssml_finetuning}). During this process, the \textit{StorySync} toolkit evolved to include granular SSML controls for prosodic attributes.

\subsubsection{Incorporating Ambient Cues}
\begin{figure}[h!]
    \centering
    \includegraphics[width=0.75\linewidth]{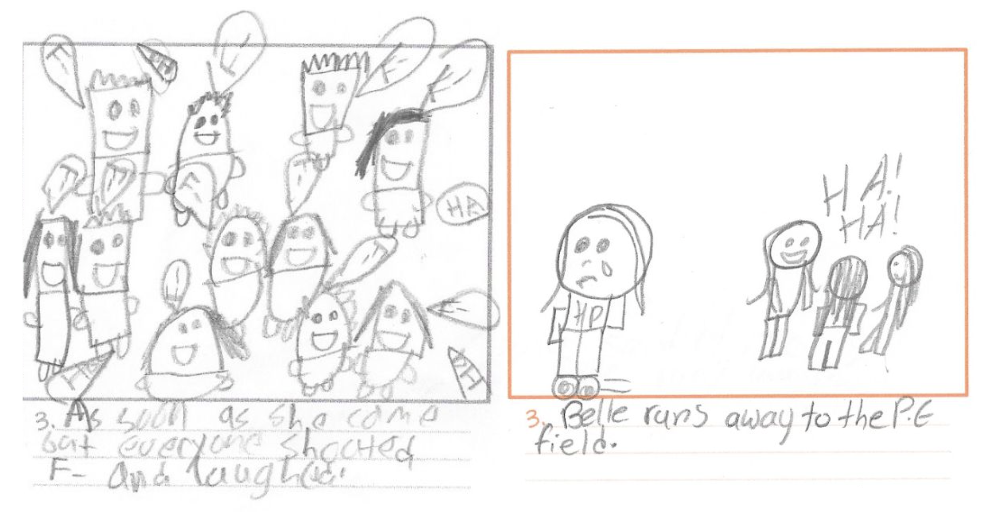}
    \caption{Children’s storyboards in \cite{sanoubari2021robots} often depicted group laughter in peer bullying scenarios, which informed the sound design of \textit{REMind} bullying scene.}
    \label{fig:groupLaughter}
\end{figure}

\textit{REMind} used ambient audio-visual cues to convey emotional subtext beyond robot's behavior. For example, to convey the social pressure in peer bullying, the dramatic scene included a background sound of group laughter, drawing inspiration from children’s storyboards (see Fig. \ref{fig:groupLaughter}). Or, to convey the affective state of the victim before \textit{AVATAR}'s intervention, we used thunderstorm lightning effects paired with a heavy rain sound-escape.

\section{Discussion}
This case study suggests that, in RMAD, aesthetics should not be understood as decorative surface features, but as part of the pedagogical infrastructure of the experience. In \textit{REMind}, the aesthetic experience of drama was necessary for supporting attention, emotional engagement, and reflective participation; yet this experience could not be carried by the robot actors alone. Because current social robots have limited capacity for conveying nuanced emotional subtext, affect had to be distributed across the broader system---including dialogue, ambient light and music, props, GUI elements, timing, and facilitation. From this perspective, the challenge of RMAD is not merely to make robots more expressive, but to choreograph a multimodal ecology in which limited robot performances can still become dramatically meaningful and emotionally resonant.

This manuscript discusses one particularly difficult subset of aesthetics in RMAD: how to create emotional subtext and affective legibility despite the expressive limitations of current robot platforms. However, this is not a comprehensive account of the aesthetics of \textit{REMind}. Other aesthetic dimensions of the experience were also important to children’s enjoyment and immersion, including imaginative pretend-play with props, ritualized actions that made the room feel responsive to the child’s behavior, relatable storytelling (designed by children for children), digital scenography, and the sustained sense of interactivity throughout the experience. Although \textit{REMind} was largely pre-scripted, participants were rarely passive observers: they were continuously engaged in \textit{doing} things---answering robots’ questions, manipulating objects, making interpretive choices, and improvising responses. Taken together, this suggest that the aesthetics of RMAD are distributed not only across robot performance, but across the entire embodied and make-believe interaction ecology of the experience.

This has two implications for HRI design. First, it highlights the value of treating robot-mediated learning experiences as integrated systems in which learning goals, narrative, aesthetics, technology, and facilitation must be intentionally aligned. Second, it suggests that human facilitation is not simply a temporary workaround for immature autonomy, but may be a constitutive part of RMAD as an experiential medium. For this reason, future work should move beyond evaluating isolated robot behaviors and instead investigate how different configurations of robot performance, scenography, and facilitation shape children's reflection and meaning-making.

\section{Conclusion}
This paper presented \textit{REMind} as a case study in Robot-Mediated Applied Drama (RMAD) and examined the challenge of sustaining the aesthetic experience of drama despite the expressive limitations of social robots. We argue that, in RMAD, aesthetics are part of the pedagogical infrastructure of the experience, and that aesthetic engagement cannot be reduced to robot expressivity alone. Instead, meaningful robot-mediated drama emerges from the coordination of robot performance with facilitation, scenography, props, timing, and participant action. We hope this case contributes to ongoing efforts in HRI to develop more performative, embodied, and dramaturgically informed uses of social robots.

\begin{credits}
\subsubsection{\ackname} We thank psychologists Dr.\ Christina Salmivalli and Dr.\ Claire Garandeau (University of Turku), and the Centre for Spectatorship and Audience Research (University of Toronto) for expertise and feedback that shaped \textit{REMind}.
\end{credits}

\bibliographystyle{splncs04}
\bibliography{references}

@book{sellers2017advanced,
    title = {{Advanced game design: a systems approach}},
    year = {2017},
    author = {Sellers, Michael},
    publisher = {Addison-Wesley Professional}
}

@misc{apple2025arkit,
    title = {{Apple Inc. : ARKit}},
    year = {2025},
    url = {developer.apple.com/augmented-reality/arkit/}
}

@article{new2007category,
    title = {{Category-specific attention for animals reflects ancestral priorities, not expertise}},
    year = {2007},
    journal = {Natl. Acad. Sci.},
    author = {New, Joshua and Cosmides, Leda and Tooby, John},
    number = {42},
    pages = {16598--603},
    volume = {104},
    publisher = {National Academy of Sciences}
}

@article{chandra2020children,
    title = {{Children teach handwriting to a social robot with different learning competencies}},
    year = {2020},
    journal = {J. Soc. Robotics},
    author = {Chandra, Shruti and Dillenbourg, Pierre and Paiva, Ana},
    number = {3},
    pages = {721--748},
    volume = {12},
    publisher = {Springer}
}

@inproceedings{sanoubari2022designing,
    title = {{Designing an Anti-Bullying Serious Game: Insights from Interviews with Teachers}},
    year = {2022},
    booktitle = {Joint Intl. Conf. on Serious Games (JCSG)},
    author = {Sanoubari, Elaheh and Cardona, John E Muñoz and Houston, Andrew and Young, James and Dautenhahn, Kerstin},
    pages = {102--121},
    organization = {Springer}
}

@article{decoursey2018embodied,
    title = {{Embodied aesthetics in drama education}},
    year = {2018},
    author = {DeCoursey, Matthew},
    publisher = {Bloomsbury Publishing}
}

@article{norman2005emotional,
    title = {{Emotional design: People and things}},
    year = {2005},
    journal = {Retrieved February},
    author = {Norman, Donald A},
    volume = {1}
}

@misc{epicgames_livelinkface,
    title = {{Epic Games: Live Link Face}},
    year = {2024},
    url = {dev.epicgames.com/documentation/en-us/unreal-engine/live-link-face-device}
}

@misc{furhat2025gesturecapture,
    title = {{Furhat Robotics: Gesture Capture}},
    year = {2025},
    url = {docs.furhat.io/gesture_capture_tool/}
}

@inproceedings{al2012furhat,
    title = {{Furhat: a back-projected human-like robot head for multiparty human-machine interaction}},
    year = {2012},
    booktitle = {Cognitive Behavioural Systems},
    author = {Al Moubayed, Samer and Beskow, Jonas and Skantze, Gabriel and Granstr{\"{o}}m, Björn},
    pages = {114--130},
    organization = {Springer}
}

@inproceedings{hicks2019juicy,
    title = {{Juicy game design: Understanding the impact of visual embellishments on player experience}},
    year = {2019},
    booktitle = {The annual symposium on computer-human interaction in play},
    author = {Hicks, Kieran and Gerling, Kathrin and Dickinson, Patrick and Vanden Abeele, Vero},
    pages = {185--197}
}

@article{salmivalli2012kiva,
    title = {{KiVa antibullying program: Overview of evaluation studies based on a randomized controlled trial and national rollout in Finland}},
    year = {2012},
    journal = {International Journal of Conflict and Violence},
    author = {Salmivalli, Christina and Poskiparta, Elisa},
    number = {2},
    pages = {293},
    volume = {6},
    publisher = {International Journal of Conflict and Violence}
}

@article{matheus2025long,
    title = {{Long-term interactions with social robots: Trends, insights, and recommendations}},
    year = {2025},
    journal = {ACM Transactions on Human-Robot Interaction},
    author = {Matheus, Kayla and Ramnauth, Rebecca and Scassellati, Brian and Salomons, Nicole},
    number = {3},
    pages = {1--42},
    volume = {14},
    publisher = {ACM New York, NY}
}

@inproceedings{hunicke2004mda,
    title = {{MDA: A formal approach to game design and game research}},
    year = {2004},
    booktitle = {AAAI Worksh. on Challenges in Game AI},
    author = {Hunicke, Robin and LeBlanc, Marc and Zubek, Robert},
    pages = {1722}
}

@inproceedings{sanoubari2021robots,
    title = {{Robots, Bullies and Stories: A Remote Co-design Study with Children}},
    year = {2021},
    booktitle = {Interaction Design and Children},
    author = {Sanoubari, Elaheh and Mu{\~{n}}oz Cardona, John Edison and Mahdi, Hamza and Young, James E and Houston, Andrew and Dautenhahn, Kerstin},
    pages = {171--182}
}

@incollection{breazeal2016social,
    title = {{Social robotics}},
    year = {2016},
    booktitle = {Springer handbook of robotics},
    author = {Breazeal, Cynthia and Dautenhahn, Kerstin and Kanda, Takayuki},
    pages = {1935--1972},
    publisher = {Springer}
}

@article{belpaeme2018social,
    title = {{Social robots for education: A review}},
    year = {2018},
    journal = {Science robotics},
    author = {Belpaeme, Tony and Kennedy, James and Ramachandran, Aditi and Scassellati, Brian and Tanaka, Fumihide},
    number = {21},
    volume = {3},
    publisher = {Science Robotics}
}

@book{schell2008art,
    title = {{The Art of Game Design: A book of lenses}},
    year = {2008},
    author = {Schell, Jesse},
    publisher = {CRC press}
}

@book{barba2005secret,
    title = {{The secret art of the performer: a dictionary of theatre anthropology}},
    year = {2005},
    author = {Barba, Eugenio and Savarese, Nicola},
    publisher = {Routledge}
}

@book{boal2000theater,
    title = {{Theater of the Oppressed}},
    year = {2000},
    author = {Boal, Augusto},
    publisher = {Pluto press}
}

@book{koster2013theory,
    title = {{Theory of fun for game design}},
    year = {2013},
    author = {Koster, Raph},
    publisher = {" O'Reilly Media, Inc."}
}

@inproceedings{sanoubari2022robot-mediated,
    title = {{Using Robot-Mediated Applied Drama to Foster Anti-Bullying Peer Support}},
    year = {2022},
    booktitle = {Robophilosophy},
    author = {Sanoubari, Elaheh and Johnson, Amanda and Munoz, John Edison and Houston, Andrew and Dautenhahn, Kerstin}
}

@inproceedings{sanoubari2024makes,
    title = {{What Makes an Educational Robot Game Fun? Framework Analysis of Children’s Design Ideas}},
    year = {2024},
    booktitle = {International Conference on Social Robotics},
    author = {Sanoubari, Elaheh and Mu{\~{n}}oz, John Edison and Yamini, Ali and Randall, Neil and Dautenhahn, Kerstin},
    pages = {40--55},
    organization = {Springer}
}

@article{riek2012wizard,
    title = {{Wizard of Oz Studies in HRI: A Systematic Review and New Reporting Guidelines}},
    year = {2012},
    journal = {Journal of Human-Robot Interaction},
    author = {Riek, Laurel},
    number = {1},
    pages = {119--136},
    volume = {1}
}
\end{document}